# Coherent entangled light generated by quantum dots in the presence of nuclear magnetic fields.

R. M. Stevenson[1*], C. L. Salter[1,2], A. Boyer de la Giroday[1,2], I. A. Farrer[2], C. A. Nicoll[2], D. A. Ritchie[2], and A. J. Shields[1]

[1]Toshiba Research Europe Limited, 208 Science Park,
Cambridge CB4 0GZ, UK.
[2]Cavendish Laboratory, University of Cambridge, JJ Thomson Avenue,
Cambridge CB3 0HE, UK
[*]Corresponding author email:mark.stevenson@crl.toshiba.co.uk

**Abstract**

A practical source of high fidelity entangled photons is desirable for quantum information applications and exploring quantum physics. Semiconductor quantum dots have recently been shown to conveniently emit entangled light when driven electrically, however the fidelity was not optimal. Here we show that the fidelity is not limited by decoherence, but by coherent interaction with nuclei. Furthermore we predict that on 100μs timescales, strongly enhanced fidelities could be achieved. This insight could allow tailoring of quantum logic to operate using quantum dots in the fault tolerant regime.

**Main Text**

Quantum logic based on linear optics [1] is emerging as a promising architecture to realise quantum computers powerful enough to solve currently intractable problems in science and engineering. One of the resources required for a quantum processor are entangled photons, which recently have been generated electrically in an entangled-light-emitting diode (ELED) incorporating a single semiconductor quantum dot [2]. The practical advantages of an electrically operated entangled light source are appealing, as they can be easily controlled, and many independent devices may be incorporated into integrated quantum optical circuits. However the entanglement fidelity of ELEDs is currently sub-optimal which can result in errors in quantum gates. Here we explore the limits behind the entanglement fidelity, to reveal interactions of excitons with nuclei as a dominant process. Furthermore we predict significantly enhanced fidelity under different measurement conditions for the same ELED device.

Previous experiments using pulsed optical excitation have already revealed a great deal about the entanglement dynamics of photon pairs emitted by quantum dots [3, 4, 5]. Effects previously considered include background light from other regions of the sample (such as the wetting layer), detector characteristics such as jitter and dark counts, and properties of the quantum dot itself such as radiative lifetimes, the fine-structure-splitting between the optically active exciton levels, exciton-spin scattering and cross-dephasing. The latter term describes the process whereby the coherence between the two intermediate photon-exciton eigenstates states is lost, without affecting the polarisation (spin) properties of the states, and was found previously to have only a weak effect [4]. In the work presented here decoherence of this type was found to be negligible so we shall disregard its effect from the start to simplify the discussion, though we will show certain coherent effects can give rise to similar apparent dephasing when averaged over time.

Models based on the above parameters cannot however account for all observed behaviour. For example, the majority of quantum dots with small splitting tend to show stronger polarisation anti-correlation in the circular basis, than polarisation correlation in the linear bases [6, 2, 7, 8, 9]. The effect is more pronounced in recent measurements of an ELED in d.c. mode, where anti-bunching persists for longer for co-circularly polarised photons [2].

The unpredicted behaviour in experiments may be attributed to weakly coupled eigenstates that are partially circularly polarised [6]. This would lead to circularly polarised photon pair states that evolve more slowly, resulting in persistent anti-bunching and stronger time-averaged correlations. The most obvious underlying coupling mechanism responsible is interaction of the exciton spin states with the atomic nuclei within the semiconductor quantum dot, which has been extensively studied in quantum dots in terms of its strength, polarisation properties, and influence on spin-storage [10, 11, 12, 13, 14]. The isotopes of gallium, indium and arsenic that make up the quantum dot are all spin active, and each randomly oriented nuclear spin interacts with the exciton states via the hyperfine interaction. For the purposes of our analysis, we adopt the convention of approximating the effect of the nuclear population as a net magnetic field $B_N$ (Overhauser field). The net nuclear field fluctuates on a timescale of ~100 µs, has random orientation, and normally distributed strength with mean zero and standard deviation of tens of mT [10, 15].

The fine-structure properties of single quantum dot excitons in the presence of external magnetic fields are strongly orientation dependent [16, 17]. When the field is applied in-plane, hybridisation of optically active with inactive states results in no change in the polarisation of the exciton eigenstates, and a weak, quadratic change in the fine-structure splitting. For our quantum dots the magnitude of the change is ~2µeVT$^{-2}$ [17], so the expected modification of the fine structure is of the order 1neV. This is far smaller than the typical splitting of entanglement-optimised quantum dots of the order 1µeV. The effect of nuclear fields in the plane can therefore be neglected for this analysis.

When applied in the normal direction, a magnetic field lifts the degeneracy between exciton states of opposing spin, resulting in a significant linear change in the fine-structure-splitting of ~200µeVT$^{-1}$ [18]. Thus we expect fluctuations of the fine-structure-splitting of the order of a few µeV due to the net nuclear field in this direction, which is comparable to the typical splitting of quantum dots. Therefore for our analysis, we must include the effect of fluctuating nuclear magnetic fields in the normal direction. In addition, diamagnetic effects are neglected as they are polarisation independent [16, 17, 18]

The effect of a normally oriented magnetic field on the energies of the upper ($X_u$) and lower ($X_l$) exciton levels are shown schematically in Figure 1(a). The horizontal axis represents the circularly polarised component of the fine-structure-splitting, which is proportional to the magnetic field. The plotted energies and depicted polarisations of the states $X_u$ and $X_l$ are solutions to the following Hamiltonian expressed in the {$X_H$, $X_V$} basis, which denote exciton states that emit horizontal (H) and vertical (V) polarised photons.

$$H = \frac{1}{2} \cdot \begin{pmatrix} -S_r & iS_c \\ -iS_c & S_r \end{pmatrix}$$

Here $S_c$ and $S_r$ denote the circular and rectilinear components of the fine-structure-splitting S between the optically active exciton eigenstates, given by $S = \sqrt{S_c^2 + S_r^2}$. Significant modification of the splitting is predicted for $S_c \geq S_r$; the polarisation of photons emitted by the exciton eigenstates are expected to be linear for $S_c \ll S_r$, circular for $S_c \gg S_r$, and elliptical for $S_c \sim S_r$, as indicated by arrows in Figure 1(a). We note that $S_c$ cannot be determined experimentally by the usual spectroscopy techniques used to resolve circular polarisation splitting, as fluctuations of $S_c$ over the course of the measurement average to zero. In contrast, $S_r$ can be determined experimentally, as limited spectral resolution results in measurement of the average photon energy projected onto each linear polarisation, which is independent of $S_c$.

The modification in fine-structure-splitting and polarisation of the exciton states by fluctuating $S_c$, introduced by the net nuclear field, consequently changes the form of the entangled photon pair state emitted by the decay of the biexciton state (XX) in a quantum dot. This is shown schematically in Figure 1(c). The decay proceeds via either $X_l$ or $X_u$, in each case emitting a pair of photons. The polarisation of each photon is shown, and in the general case are all found to be non-equivalent, elliptically polarised states. Including the phase difference acquired during the time τ spent in the intermediate entangled exciton-photon state [5], the emitted entangled photon pair state is;

$$\Psi_i = (\alpha|H_{XX}\rangle - i\beta|V_{XX}\rangle)(\alpha|H_X\rangle + i\beta|V_X\rangle)/\sqrt{2}$$
$$+ e^{iS\tau/\hbar}(\alpha|V_{XX}\rangle - i\beta|H_{XX}\rangle)(\alpha|V_X\rangle + i\beta|H_X\rangle)/\sqrt{2}$$

where weights $\alpha$ and $\beta$ are shown in Figure 1(b) and are given by the relationships $\alpha^2 + \beta^2 = 1$, and $\beta/\alpha = S_c/(S_r + S)$.

Inspection of the above equation reveals that for $S_c \sim 0$, $\beta \sim 0$ and $\Psi_i \approx (|H_{XX}\rangle|H_X\rangle)/\sqrt{2} + e^{iS\tau/\hbar}(|V_{XX}\rangle|V_X\rangle)/\sqrt{2}$, reverting to the usual form for the entangled state from a quantum dot [5]. As expected, for large $S_c$, $\beta \sim \alpha$, and entangled photons evolve in the circular basis, evident from the two photon state $\Psi_i \approx (|R_{XX}\rangle|L_X\rangle)/\sqrt{2} + e^{iS\tau/\hbar}(|L_{XX}\rangle|R_X\rangle)/\sqrt{2}$, where $|L\rangle = (|H\rangle + i|V\rangle)/\sqrt{2}$ and $|R\rangle = (|H\rangle - i|V\rangle)/\sqrt{2}$ are left and right hand circularly polarised photons.

We now describe our model used to simulate entangled photon emission from quantum dots. It is based on a system of rate equations derived from a multi-level approximation of the quantum dot, as depicted in Figure 2(a). Since the exciton basis states of the quantum dot are dependent on the fluctuating nuclear magnetic field, polarisation and/or spin is a poor choice to differentiate between states, so instead we label the four levels in terms of the number of excitons in the dot, and their coherent properties. The resulting levels are the initial coherent exciton state $X_C$, produced by emission of a previously detected coherent biexciton photon, and the incoherent ground (G), exciton ($X_S$) and biexciton (XX) states. The levels are linked by re-excitation at rate proportional to $p$ (yellow arrows), radiative decay at rates $\Gamma_{XX}$ and $\Gamma_X$ (purple arrows) and spin-scattering at rate $\Gamma_S$ (green arrow).

The rate equations are constructed and solved using standard methods, and the density matrix that describes the intensity and polarisation of the two-photon state $\rho$ is computed as follows;

$$\rho(S_r, S_c, \tau) \propto k(X_C(\tau)\rho_e(S_r, S_c, \tau) + X_S(\tau)\rho_m) + (1-k)\rho_m$$

where $\tau$ is the time delay between emission of the biexciton and exciton photons, k is the fraction of light originating exclusively from the dot, $\rho_e$ is the density matrix for entangled light $|\Psi\rangle\langle\Psi|$, and $\rho_m$ the density matrix for maximally mixed light equal to the identity matrix $I_4/4$. Similar methods are used to construct the density matrix for negative time delays, which correspond to emission of an XX photon after detection of an X photon.

To account for fluctuations in the nuclear magnetic field, a weighted average of $\rho$ is numerically obtained from a normally distributed $S_c$ with standard deviation σ and mean zero, as $S_c$ is proportional to the normally distributed nuclear magnetic field. The resulting distribution of splitting magnitudes $|S|/\sigma$ is shown in Figure 2(b), for different ratios of $S_r/\sigma$. For $S_r/\sigma=0$, the distribution of $|S|$ is Gaussian with standard deviation σ, as it is defined solely by the effect of the nuclear magnetic field. For increasing $S_r/\sigma$, the form of the distribution becomes more exponential-like, accompanied by distinct narrowing. This illustrates that for large $S_r/\sigma$, the magnitude of the splitting is almost constant so nuclear magnetic field fluctuations are unimportant.

The model was used to fit the second-order cross correlation measurements presented in [2], for a d.c. electrically excited quantum dot. The d.c. injection scheme is ideally suited for this analysis as the excitation rate is independent of time, and correlations are not as strongly dominated by the radiative decay time compared to pulsed excitation. However the model and conclusions discussed here are equally relevant to pulsed and/or optical excitation. Second-order cross correlations were calculated as a function of time delay τ, by projection of the calculated density matrix onto the corresponding polarised photon pair states. Finally the traces were convolved with a Gaussian approximation of the actual detector response functions to account for finite detector resolution. The corresponding experimental and fitted correlations are shown in Figure 3 for photon pairs measured in the rectilinear (a), diagonal (b) and circular (c) polarisation bases. Excellent qualitative agreement is observed, revealing nuclear spin effects to be responsible for much of the characteristic shape of photon correlations.

The fitted parameters were k=0.866, σ=2.47μeV, and $\Gamma_S$=0, where k represents the fraction of photon pairs originating from the dot and is found to be similar to previous measurements [5]. All other parameters were experimentally measured. The fact that no spin-scattering term is required is surprising, as it suggests that photon pair emission from the quantum dot does not suffer dephasing and is fully coherent.

The fitted standard deviation σ of the circular component of the fine structure splitting $S_c$ is equivalent to the standard deviation of the energy fluctuations caused by the nuclear spins $\mu_B g_e \Delta B_N$, where $g_e$ is the electron g factor. Our quantum dots are expected to be quite small due to their high emission energy, and consisting of ~$10^4$ atoms[19]. For dots of this size $\mu_B g_e \Delta B_N$ is ~1.5 µeV in GaAs and ~3.4 µeV in InAs, the latter stronger due to the larger value for the indium nuclear spin (9/2) compared to gallium (3/2)[11, 20, 21]. Thus the extracted value of σ=2.47µeV is consistent with a small InAs quantum dot intermixed with GaAs, as expected in our samples.

Quantitative agreement is assessed using the degree of polarisation, defined as the difference between co- and cross-polarised correlations, divided by their sum. Simulations result in optimum values of 0.60, 0.59 and -0.77 measured for the rectilinear, diagonal, and circular polarisation bases, which are in very good agreement with the corresponding experimentally measured values of 0.63±0.05, 0.57±0.06 and -0.77±0.04 respectively.

The physical effect of nuclear fields can clearly be seen in the correlation traces. For dots with very small splitting such as this (measured to be 0.4 µeV), the quantum dots have strongly circularly polarised eigenstates due to the nuclear magnetic field and thus oscillations of $g^{(2)}$ with τ are expected in the rectilinear and diagonal bases as the phase between $X_L$ and $X_R$ evolves. The frequency of these oscillations is proportional to S, thus averaging over the distribution of S results in an additional damping term, which is Gaussian in form and of width $\hbar/\sigma$~270ps, originating from the Fourier transform of the normally distributed S (see Figure 2(b)). This results in faster convergence of the correlated and anti-correlated traces in the rectilinear and diagonal bases, as observed experimentally and theoretically.

The finite temporal resolution of our measurement system means that the same damping of coherent oscillations effect reduces the maximum and minimum observable correlations due to averaging. This explains why the overall degree of correlation is stronger in the circular basis, common for dots with small splitting ($S_r$). Similarly, the finite temporal resolution also results in a reduction of the overall degree of correlation in the circular and diagonal bases for dots with non-zero $S_r$, Thus whether the polarisation correlation measured in the circular basis dominates over that measured in the linear basis[6, 2, 7, 9], or vice versa[22, 23, 24, 25, 26], depends on the nuclear spin configuration and splitting $S_r$ of the quantum dot, and the measurement technique employed.

Combining the above effects of finite $S_r$, $S_c$, and timing resolution can result in a largely uncorrelated contribution to the two-photon density matrix, which could form a component of the polarisation uncorrelated fraction of photon pairs always observed in entangled photon pair emission from dots with small $S_r$ (see e.g. refs [3, 24, 25, 26, 7]). Our conclusions may therefore offer a coherent mechanism to explain the apparent spin-scattering or depolarisation of the exciton spin states, observed in previous experiments.

It is well known that suppression of the fine-structure-splitting S is advantageous for many entangled photon experiments and applications, in order to minimise the phase evolution between orthogonally polarised photon pair states. Thus similarly to the reduction of $S_r$, it may be beneficial to supress the hyperfine interaction and hence $S_c$. This is commonly achieved by applying external magnetic fields[27, 28], or polarizing the nuclear spin population[19, 13]. In addition, the dependence of hyperfine interaction strength on the size and alloy of the quantum dot suggests using larger, GaAs rich quantum dots, will supress effects attributed to nuclear fields.

We stress that the timescale for fluctuations of nuclear fields are of the order ~100μs [10] up to several seconds in the presence of external fields [28], so although they can lead to apparent decoherence on timescales typical of present experiments, they are effectively static over the time required to do a quantum gate operation, and thus do not cause decoherence. This fact has been demonstrated numerous times by the use of echo techniques which effectively cancel the effect of the nuclear field experienced in each individual cycle [12, 28]. Therefore the entangled state emitted by a quantum dot is expected to be stable for ~$10^4 - 10^8$ cycles, when operating at 100MHz, which in principle is sufficient for many applications, negating the need to control the hyperfine interaction at all. This is true even for finite fine-structure-splitting S, as the time-evolving entangled state can be temporally resolved.

It is interesting to estimate the entanglement properties of the ELED accessible to future applications. For this purpose, we begin with the measured and calculated time-averaged fidelity of the entangled light as a function of the time delay between photons, plotted in Figure 3(d). The measured (red) and calculated (black) curves match very well, with peak fidelities of 0.71±0.02 and 0.73 in agreement (within error). Next, we remove the effect of the fluctuating nuclear field, (dark grey curve), which improves the peak fidelity significantly from 0.73 to 0.82. In addition, the time over which entangled light is emitted, defined by fidelities>0.5, is significantly extended from 1.0 ns to 1.8 ns, as the damping effect described above is removed.

Finally, we remove the effect of detector jitter (light grey curve) to reveal an underlying fidelity of 0.9 for this particular device, limited only by the dark counts and background light, which define the parameter k. Attempts to increase k by better spectral and spatial filtering of quantum dot light could improve the fidelity further.

In conclusion, we have shown that the fidelity of entangled photons emitted by quantum dots is limited by coherent interactions of excitons with nuclei, presence of background light, and detector jitter, and not by any intrinsic decoherence. Improved measurement techniques are expected to eliminate the time-averaging effects that lead to apparent reduction in the entanglement fidelity, and allow enhanced fidelities to be measured for the same quantum dot source. The high inherent entanglement fidelity in quantum dot based ELEDs combined with the practical advantages of an electrically driven semiconductor device make such entangled light sources very promising for future quantum information applications.

**Acknowledgements**

We would like to acknowledge funding for this work from the EPSRC, QIPIRC, QAP, Q-ESSENCE, NanoSci-ERA and Spin-Optronics ITN. We thank K. Cooper for advice and support in device fabrication.

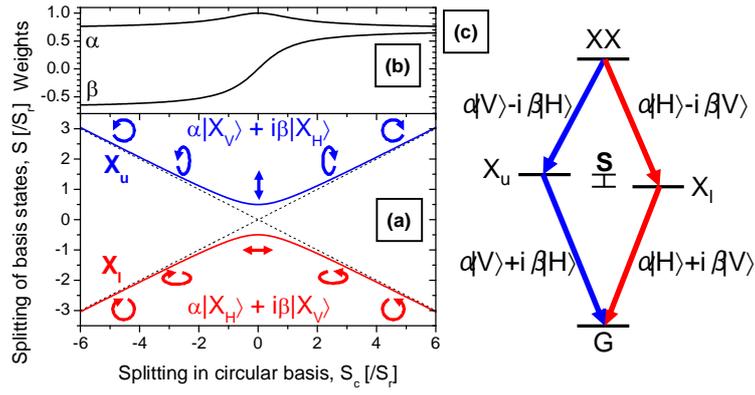

Figure 1. Schematic showing effect of normally oriented magnetic fields on the exciton states of a quantum dot. (a) Upper ($X_u$) and lower ($X_l$) energy exciton eigenstates of a quantum dot, as a function of the circularly polarised component $S_c$ of the fine-structure-splitting S. Units are the rectilinearly polarised component $S_r$ of S. The polarisation of light emitted by the excitons is indicated by arrows. (b) Weights α and β of the linearly polarised components of the eigenstates shown in (a). (c) Schematic of biexciton (XX) decay in a quantum dot via one of the exciton eigenstates $X_u$ and $X_l$ to the ground state G. Arrows indicate photon emission with polarisation as marked.

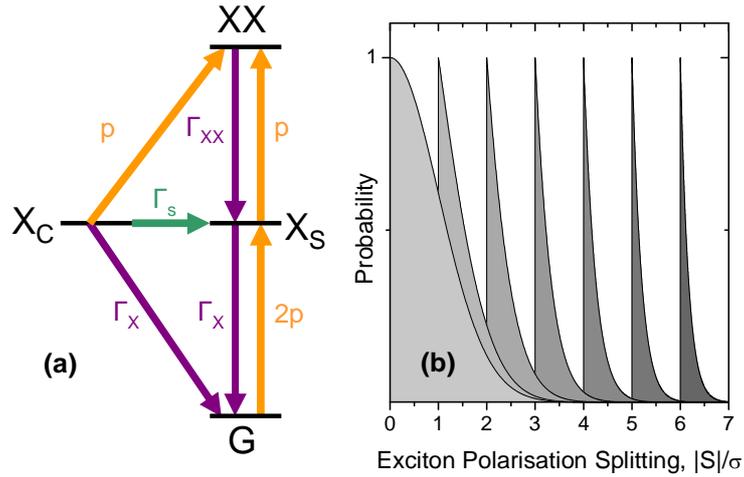

Figure 2. Schematic of calculations. (a) Four level model includes biexciton state (XX), coherent and mixed exciton states ($X_c$ and $X_s$), and ground state (G). Orange arrows indicate excitation by electrical current at rate p, and purple arrows indicate radiative decay from XX and X states at rates $\Gamma_{XX}$ and $\Gamma_X$. Green arrow indicates a potential decoherence process in the form of exciton spin-scattering at rate $\Gamma_S$. (b) Distribution of the magnitude of fine-structure-splitting S expected during an experiment. The rectilinearly polarised component $S_r$ of S increases from 0 to $6\sigma$ between lightest and darkest grey curves. The unit $\sigma$ is the standard deviation of the fluctuating circularly polarised component $S_c$ of S.

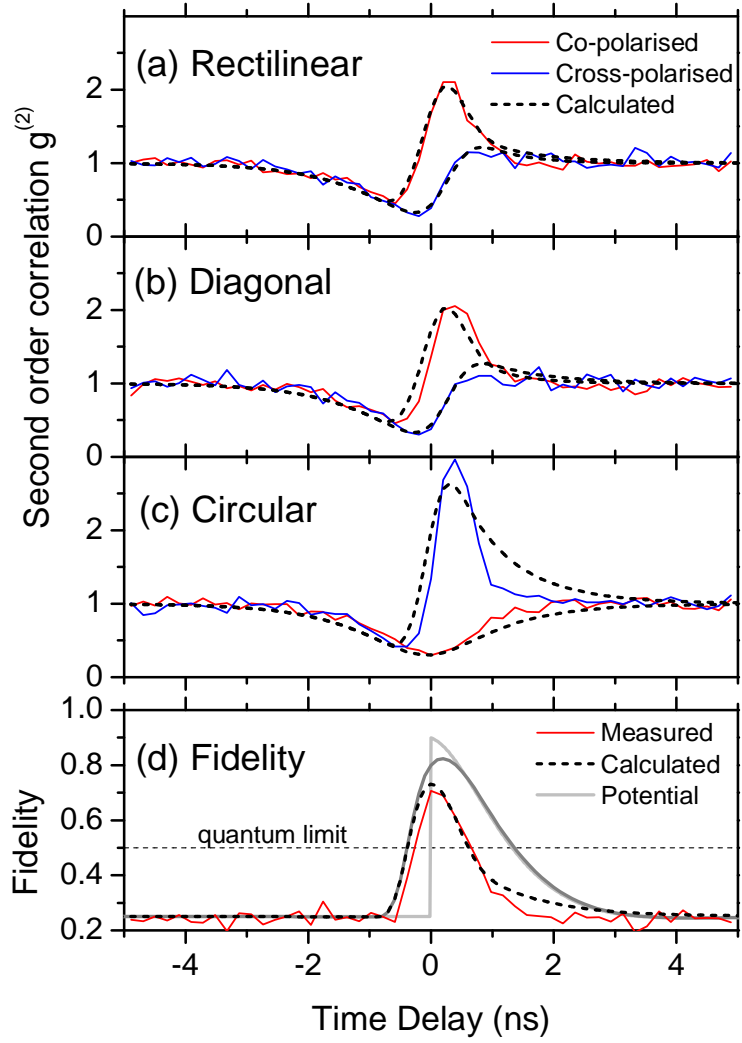

Figure 3. Comparison between experimental and simulated entangled photon emission from an ELED. (a-c) Second order correlation measurements as a function of the delay between XX and X photons, $g^{(2)}(\tau)$. Red and blue curves are co- and cross polarised experimental measurements from [2], performed in the (a) rectilinear, (b) diagonal, and (c) circular polarisation bases. Dashed black curves are corresponding simulated behaviour with parameters as described in the text. (d) Fidelity of photon pair emission with the Bell state $(|HH\rangle+|VV\rangle)/\sqrt{2}$, as a function of the time delay. Red curve is experimental result from [2], and dashed black curve is corresponding simulation. Predicted behaviour for short measurements is shown by dark grey, and additionally with fast detectors by light grey lines.